\documentclass[sigconf,nonacm]{acmart}
\usepackage{hyperref}
\usepackage{graphicx}
\usepackage{subcaption}
\usepackage{listings}
\usepackage{booktabs}
\usepackage{caption}
\usepackage{xcolor}
\usepackage{framed}
\newenvironment{conclusionbox}{%
    \MakeFramed{\advance\hsize-\width\FrameRestore}%
    \noindent\ignorespaces
}{\endMakeFramed}

\begin{document}

\title{AI-Generated Smells: An Analysis of Code and Architecture in LLM- and Agent-Driven Development}

\author{Yue Cai Zhu, Nikolaos Tsantalis, Peter C. Rigby}
\affiliation{%
  \institution{Department of Computer Science \\and Software Engineering}
    \institution{Concordia University}
  \city{Montr{\'e}al, Qu{\'e}bec}
  \country{Canada}
  }

\begin{abstract}
The promise of Large Language Models (LLMs) in automated software engineering is often measured by functional correctness, overlooking the critical issue of long-term maintainability. This paper presents a systematic audit of technical debt in AI-generated software, revealing that AI does not eliminate flaws but rather introduces a distinct "machine signature" of defects. Our multi-scale analysis—spanning single-file algorithmic tasks and complex, agent-generated systems—identifies a fundamental Reasoning-Complexity Trade-off: as models become more capable, they generate increasingly bloated and coupled code. This architectural decay is so pronounced that we establish a Volume-Quality Inverse Law, where code volume is a near-perfect predictor of structural degradation. Crucially, we demonstrate that neither functional correctness nor detailed prompting mitigates this decay. These findings challenge the current paradigm of prompt-driven generation, reframing the central problem of AI-based software engineering from one of code generation to one of architectural complexity management. We conclude that future progress depends on equipping agents with explicit architectural foresight to ensure the software they build is not just functional, but also maintainable.

\end{abstract}

\maketitle

\section{Introduction}
The advent of Large Language Models (LLMs) has fundamentally altered the landscape of automated software engineering, enabling agents to autonomously synthesize functional software ranging from single-file scripts to complex, multi-file architectures. While multi-agent frameworks show increasing proficiency in handling intricate logic, the efficacy of these systems is often evaluated solely on functional correctness. This narrow focus risks overlooking critical concerns regarding code quality, specifically the presence of code smells—structural characteristics that, while functionally valid, indicate deeper problems in design, readability, and maintainability.

Previous research has established that human developers frequently introduce code smells under tight constraints, accumulating technical debt that degrades maintainability. Conversely, recent studies suggest that LLMs may possess the capacity to generate code that is statistically more maintainable than human-written code. For instance, Santa et al.~\cite{santa2025llm} demonstrated that LLMs produced fewer issues detected by SonarQube, resulting in a lower estimated time for remediation. However, these findings often aggregate superficial code linting issues with deeper structural code smells. Furthermore, the extent to which these positive trends persist within autonomous agents constructing complex systems remains underexplored. Consequently, there is a pressing need to specifically audit agent-generated code for architectural design flaws.

This study presents a systematic evaluation of code smells in software produced by single-agent and multi-agent systems. We specifically analyze how the transition from simple tasks to complex, multi-file architectures influences the density and distribution of these smells. To guide this investigation, we pose the following research questions:

\textbf{RQ1: To what extent do code-generating agents introduce code smells in standard coding tasks?}
This question aims to establish a baseline for AI code quality. By quantifying the density of code smells in standard benchmarks, we determine whether LLM-based agents inherently prioritize functional output at the expense of code hygiene and readability.

\textbf{RQ2: What constitutes the taxonomy of code smells most frequently exhibited by code-generating agents?}
Not all technical debt is identical. By categorizing specific types of smells (e.g., Message Chain, Too Many Branches, or Hub-like Dependency), we identify the specific coding "habits" of LLMs. This taxonomy is essential for understanding how agents handle abstraction and responsibility distribution.

\textbf{RQ3: How does the complexity of the target architecture impact the prevalence of code smells in agent-generated code?}
As agents transition from generating isolated scripts to synthesizing multi-module systems, context management becomes increasingly difficult. We investigate whether this increased architectural complexity correlates with a degradation in design patterns, and whether specific prompt strategies can mitigate this decay.

To answer the aforementioned research questions, we conduct two experiments. In the first experiment, we ask different LLMs to solve coding questions sampled from the CodeContest\cite{li2022competition} dataset. In the second experiment, we create a set of product requirements to benchmark the code quality of the software repositories generated by MetaGPT\cite{hong2024metagpt}.

\textbf{Summary of Findings and Contributions.} Our empirical results reveal that AI agents do not merely replicate human errors but exhibit a distinct "machine signature" of technical debt.
In algorithmic tasks (RQ1), we observe a "Reasoning-Complexity Trade-off": while humans struggle with state encapsulation (\textit{Temporal Fields}), "smarter" LLMs (e.g., Qwen-480b) inadvertently increase method bloat (\textit{Long Method}) as they attempt to handle complex logic within single procedural blocks.
As we scale to complex architectures (RQ2), a taxonomic shift occurs: the method-level bloat is replaced by "God Class" syndromes (\textit{Too Many Branches}) and Redundant Implementation (\textit{Potential Improper API Usage}).
Crucially, our analysis of architectural complexity (RQ3) establishes a "Volume-Quality Inverse Law": code volume (TLoC) acts as a near-perfect predictor of architectural decay. We further find that functional correctness is decoupled from quality—running code is just as likely to be structurally flawed as failed code—and that increasing requirement specificity in prompts fails to mitigate this degradation.

Based on these insights, this paper makes the following contributions:

\begin{conclusionbox}

\textbf{1. We present a comparative analysis of technical debt in human vs. LLM-generated code,} identifying a fundamental "habit" of LLMs: a preference for high-coupling strategies—either through monolithic methods (\textbf{Long Method}) in simple tasks or tight module coupling (\textbf{Unstable Dependency}) in complex systems—to achieve functional correctness at the cost of maintainability.

\textbf{2. We conduct the first longitudinal study of "architectural rot" in autonomous software agents,} demonstrating that as system complexity scales, functionally correct code degenerates into unmaintainable structures. Our statistical analysis confirms that this decay is not mitigated by advanced prompting techniques like few-shot examples. 

\textbf{3. We establish a taxonomy of AI-specific code smells and introduce the concept of the "Modular Mirage,"} where agents achieve superficial structural modularity (file separation) but fail to create semantic cohesion. This reframes the central challenge of LLM-based software engineering from *code generation* to *complexity management*.
\end{conclusionbox}

\section{Background and Related Work}
Recent advancements in Large Language Models (LLMs) have catalyzed significant research into AI-driven software engineering. A substantial body of work has explored the utility of LLMs for tasks such as code smell localization~\cite{batole2025llm} and automated software refactoring~\cite{naik2024enhancing, wadhwa2024core, cordeiro2024empirical,liu2025exploring}. Similarly, the emergence of Multi-Agent Systems (MAS) has prompted investigations into agentic refactoring workflows~\cite{siddeeq2025llm, pucho2025refactoring,rajendran2025multi,xu2025mantra}. Despite this progress, there remains a paucity of research dedicated to analyzing the intrinsic quality and maintainability of the code generated by these autonomous agents. While agents are increasingly proficient at producing functional code, their tendency to introduce technical debt remains underexplored. This work aims to bridge this gap. To contextualize our empirical study, we first define the scope of autonomous code-generating agents.

\subsection{Autonomous Code-Generating Agents}
The landscape of software engineering has been transformed by the proliferation of autonomous code generation technologies~\cite{dong2025survey, he2025llm}. In the context of this research, we define \textit{autonomous code-generating agents} broadly to encompass a spectrum of workflows, ranging from direct interactions with Large Language Model (LLM) APIs to sophisticated multi-agent frameworks. This definition includes both atomic workflows—such as prompting a model like Gemini to generate a single function—and complex orchestrations where multiple agents collaborate to synthesize entire software systems. Both paradigms rely on the fundamental capability of generative AI to produce executable code without continuous human intervention. Adopting this broad definition allows us to investigate the quality of code generated across the full spectrum of autonomous development tools.

\subsection{LLM-Based Code Generation and Maintainability}
Large Language Models have demonstrated remarkable proficiency in code synthesis, achieving high performance on functional correctness benchmarks such as HumanEval and MBPP~\cite{zheng2023survey,chen2021evaluating,zhang2023multilingual}. However, functional correctness does not guarantee code quality. Recent literature has begun to scrutinize the maintainability of AI-generated code. Notably, Santa et al.~\cite{santa2025llm} reported that LLMs are capable of generating code that is statistically more maintainable than that of human developers. Their evaluation relied on metrics provided by SonarQube~\cite{campbell2013sonarqube}, specifically the count of detected issues and the estimated technical debt remediation time.

We contend, however, that a critical limitation in existing studies, including the work of Santa et al.~\cite{santa2025llm}, is the conflation of different categories of software defects. The "issues" detected by automated tools like SonarQube often aggregate superficial code linting violations (e.g., whitespace inconsistencies, missing docstrings) with deeper, structural code smells. While LLMs excel at adhering to syntax and formatting rules—thereby minimizing linting errors—this does not necessarily imply architectural soundness. Furthermore, prior research has predominantly focused on single-file solutions generated for isolated coding problems. Our work aims to address this limitation by specifically isolating structural code smells from superficial linting errors and extending the analysis to large-scale software artifacts generated by multi-agent systems.

\subsection{Multi-Agent Systems and MetaGPT}
To address the limitations inherent in single-model generation—such as context loss, hallucination, and a lack of self-correction—researchers have introduced Multi-Agent Systems (MAS). These frameworks assign specific roles (e.g., Architect, Engineer, QA Tester) to distinct agent instances, enabling collaborative problem-solving. MAS is capable to generate larger scale software artifacts than single-model generation.

In this study, we utilize MetaGPT~\cite{hong2024metagpt} as our primary multi-agent framework in experiment 2 to generate full-scale software solutions. MetaGPT distinguishes itself by encoding Standard Operating Procedures (SOPs) directly into the agent workflow, effectively mimicking a real-world software development company. By structuring communication and enforcing role-specific constraints, MetaGPT aims to produce complex, multi-file software projects that are theoretically more robust than those produced by a single pass of an LLM. We investigate whether this structured collaboration effectively mitigates the introduction of complex design smells in multi-module architectures.

\subsection{Code Smell Detection with PyExamine}
To systematically evaluate the structural quality of generated code, we employ PyExamine, a static analysis tool designed to detect code smells at the code, structure, and architectural levels~\cite{shivashankar2025pyexamine}. PyExamine identifies specific patterns that indicate violations of fundamental software design principles. We focus on the following set of code smells detected by the tool.

\paragraph{Code Level Smell:} Code level smells are implementation artifacts that violate coding best practices within a local scope (intra-procedural). They typically indicate that a specific piece of logic is overly complex, difficult to read, or poorly formatted, making it hard for other developers to understand or modify that specific function without introducing bugs~\cite{mo2019architecture}.

\paragraph{Structural Code Smell:}In contrast to code-level smells which are localized to individual methods, structural code smells manifest at the class or object level, reflecting deeper flaws in the system's object-oriented design. These smells arise from the improper organization of relationships between entities, often signaling violations of fundamental principles such as encapsulation, coupling, and cohesion. While code-level issues primarily affect the readability of specific logic blocks, structural smells degrade the software's maintainability by making classes overly complicated or functions overly coupling with each other. Consequently, the presence of structural smells indicates a rigid design that is difficult to extend or refactor without causing ripple effects throughout the code base~\cite{mo2019architecture}.

\paragraph{Architectural code smells:} Architectural code smells are defined as recurring architectural design decisions that negatively impact the internal quality, understandability, and maintainability of a software system. While code level smells (e.g., LM) manifest within specific implementation blocks and structural code smells characterize improper coupling between individual classes, architectural smells operate at the highest level of abstraction. They describe problematic dependencies and relationships between large-scale system components, packages, or namespaces. Common examples, such as Cyclic Dependencies or God Objects, represent a divergence between the intended architecture and the actual implementation. Consequently, whereas code and structural smells can often be mitigated through localized refactoring, architectural smells typically span the system’s macroscopic structure, necessitating more invasive and costly remediation efforts to address technical debt~\cite{mo2019architecture}.

While code smells are well-studied in human-written software, the design habits of autonomous coding agents remain largely unexplored. Most current research focuses on whether AI-generated code functions correctly, often overlooking deeper structural issues. This leaves a significant gap in our understanding of how agents manage complexity and technical debt in larger projects. To address this, we conducted a comparative study to analyze the quality of code produced by both single and multi-agent systems. The following section details our methodology and explains how we detected and measured these architectural flaws.

\section{Methodology}
To comprehensively evaluate the code quality and structural habits of autonomous agents, we designed a two-phase experimental methodology. This dual approach allows us to investigate code smells across two distinct dimensions: algorithmic logic and architectural design. In the first phase (Experiment I), we utilize the CodeContest dataset to analyze how individual Large Language Models (LLMs) handle isolated, algorithmic coding tasks. In the second phase (Experiment II), we expand the scope to multi-agent systems using MetaGPT, challenging the agents to generate complete, multi-file software repositories based on specific product requirements. By contrasting these two scenarios, we aim to isolate the impact of system complexity on the accumulation of technical debt. The remainder of this section details the experimental setup, the data generation process, and the specific static analysis tools employed for smell detection.

\subsection{Experiment I Setup}
This experiment evaluates the code quality of LLM-generated solutions across varying levels of prompt guidance. We generated Python solutions for a curated set of algorithmic problems under two distinct conditions: \textit{Zero-Shot} and \textit{Few-Shot}. This design mirrors the methodology of Santa et al.~\cite{santa2025llm}, with the modification that we focus exclusively on off-the-shelf models rather than fine-tuned variants.

The \textit{Zero-Shot} condition serves as a baseline, instructing the model to solve the problem with minimal context. The \textit{Few-Shot} condition introduces structured guidance, including coding specifications, best practices, and boilerplate examples. To quantify the quality of the output, we employed PyExamine~\cite{shivashankar2025pyexamine}, a static analysis tool capable of detecting a wide range of Python-specific code smells. By comparing the density and distribution of smells across these two conditions, we aim to determine whether explicit prompt engineering effectively mitigates the introduction of technical debt in algorithmic code generation.

\subsubsection{LLMs In The Analysis}
To ensure the generalizability of our findings, we selected a diverse set of Large Language Models (LLMs) spanning different architectures (Dense vs. Mixture-of-Experts), access types (Proprietary vs. Open Source), and parameter scales. The selected models are:

Gemini 2.5 Pro: A state-of-the-art, proprietary flagship model from Google known for its advanced reasoning and coding capabilities. It is designed to handle complex problems and can process information from various formats, including entire code repositories

Llama 3.3:70b: A high-performance 70-billion-parameter open-source model from Meta. This model is recognized for its efficiency, offering performance comparable to much larger models, and features improvements in math and coding.

deepseek-coder-v2:16b: An open-source, code-specialized Mixture-of-Experts (MoE) model that is highly efficient, with 16 billion total parameters (2.4 billion active). It is designed to rival the performance of closed-source models on code-specific tasks and supports over 338 programming languages.256

qwen3-coder:30b and qwen3-coder:480b: A series of agentic coding models from Alibaba's Qwen team, built on a Mixture-of-Experts (MoE) architecture. We include two versions to analyze the impact of scale: a 30B parameter model (3.3B active) and a large-scale 480B parameter model (35B active). Both are state-of-the-art open models designed for exceptional performance in coding and agentic tasks.

\subsubsection{Data Curation}
We utilized the CodeContest dataset~\cite{li2022competition}, a comprehensive benchmark comprising competitive programming problems aggregated from various platforms. To create a representative evaluation set for our experiments, we randomly sampled 90 unique coding problems from the dataset. These problems encompass a diverse range of algorithmic challenges and logic requirements, providing a broad basis for evaluating the code generation capabilities of the selected LLMs. The code to generate the same random sample is provided in the replication package~\cite{ReplicationPackage}.

\subsubsection{The Prompts}

We applied two distinct prompting strategies to generate solutions for the selected problems:

\paragraph{Condition A: Zero-Shot (Baseline)}
This condition replicates the prompt structure used by Santa et al.~\cite{santa2025llm}, providing the model with the problem description and a minimal instruction to generate a solution.
\begin{lstlisting}[caption=Zero-Shot Prompt]
Generate a Python code (only one solution) to solve this question.
\end{lstlisting}

\paragraph{Condition B: Few-Shot (Structured)}
To assess the impact of context on code hygiene, the second condition utilizes a structured prompt. This prompt explicitly defines coding standards, requests adherence to best practices, and provides a boilerplate example to guide the model's architectural decisions.
\begin{lstlisting}[caption=Few-Shot Prompt]
AGENT CONFIGURATION: Python Programmer

[ROLE]
You are a highly specialized code generation agent. \\
Your sole purpose is to solve programming problems by writing Python code.

[OUTPUT SPECIFICATION]
- The response MUST be a single, raw Python code block.
- DO NOT include any text, markdown formatting,\\
  comments, or explanations before or after the code.
- The code MUST be runnable as-is.

[ENTRY POINT DEFINITION]
- All code MUST be encapsulated within the following boilerplate.
  The core logic resides inside the `solution` method.
- The `input` and `output` types are fixed to `str`.
  Type conversions must be handled internally.

[BOILERPLATE]
def solution(self, input: str) -> str:
    # --- Your implementation begins here ---
\end{lstlisting}

\subsubsection{Code Smell Detection and Data Analysis}
To perform a quantitative and objective assessment of the code quality of the agent-generated code, we implemented a systematic code smell detection process. 

\paragraph{Tool Selection: PyExame} 
For this task, we leveraged PyExamine\cite{shivashankar2025pyexamine}, a robust and un-opinionated static analysis tool specifically designed for detecting code smells in Python source code. PyExamine was chosen for several key reasons:
\begin{itemize}
    \item Comprehensive Detection: It supports a wide array of well-documented code smells, from simple anti-patterns like Long Method and Large Class to more complex design issues.
    \item Static Analysis: As a static analysis tool, PyExamine examines the source code without executing it. This is crucial for our experiment, as it allows for a safe and consistent analysis of all generated code, regardless of its functional correctness or dependencies.
    \item Objectivity and Repeatability: By automating the detection process with a standardized tool, we ensure that the evaluation is objective, consistent, and repeatable across all generated solutions, eliminating the subjectivity of manual code reviews.
\end{itemize}

\paragraph{Detection and Analysis Process}
Our code smell detection workflow was executed as follows for every solution generated by each LLM:
\begin{enumerate}
    \item Code Isolation: The Python code generated by the LLM in response to a problem prompt was saved as an individual .py source file.
    \item Automated Execution: A script was used to systematically run PyExame on each of the generated source files. The tool parsed the code to apply its detection heuristics and rules.
    \item Data Extraction: For each analysis, PyExame produced a structured report detailing every code smell instance it identified. This report included the specific type of smell (e.g., Long Method, Feature Envy), its precise location within the code (line number), and associated metrics.
    \item Aggregation and Quantification: The raw data from these reports was aggregated to build a quantitative profile for each solution. The primary metrics we collected were the total count of code smells per solution and the frequency distribution of each smell type.
\end{enumerate}

To establish a comparative baseline aligned with the methodology of Santa et al.~\cite{santa2025llm}, we also subjected human-authored solutions to the same smell detection pipeline. For each of the 90 problems in our dataset, we randomly sampled a verified single Python submission to serve as a control. We then aggregated the metrics from this human baseline alongside the results from each LLM and prompting strategy. This data-driven framework facilitates a rigorous comparative analysis, enabling us to quantify the quality gap between human and machine-generated code and to isolate the specific impact of structured prompting on software maintainability.

\subsection{Experiment II Setup}
While Experiment I focuses on algorithmic correctness, Experiment II evaluates the capacity of multi-agent systems to maintain architectural integrity in full-scale software projects. We utilized MetaGPT~\cite{hong2024metagpt}, a multi-agent framework, to synthesize complete Python software repositories based on high-level product requirements.

Unlike the first experiment, which varied prompting strategies (Zero-Shot vs. Few-Shot), this experiment maintains a consistent prompting approach while varying the complexity of the requirements. Each prompt instructs the multi-agent system to construct a self-contained application, explicitly requiring the use of Object-Oriented Programming (OOP) principles and modular file structures. Following generation, we collected the complete directory structures—including source code, configuration files, and documentation—and analyzed them using PyExamine. In this context, PyExamine is utilized to detect not only low-level code smells but also high-level architectural anomalies, such as improper coupling across modules.

To facilitate future research, we have compiled the generated repositories and the corresponding prompt set into a benchmarking dataset, which will be released alongside this paper.

\subsubsection{Test Scenarios and Complexity}
For this evaluation, we employed the \textbf{Qwen-Coder 480B} model as the underlying LLM for the MetaGPT agents. We designed five distinct application scenarios to serve as test cases, spanning domains from interactive gaming to enterprise-grade data management systems.

Preliminary trials indicated that without explicit architectural constraints, agents frequently defaulted to creating monolithic scripts or "God Classes"—a severe code smell. To mitigate this, we engineered the prompts to strictly enforce modularization and separation of concerns. Furthermore, to simulate the realistic lifecycle of software evolution, we adopted an incremental specification strategy. For each of the five scenarios, we developed a sequence of four prompts with progressively increasing complexity. This longitudinal approach allows us to observe how the agent's architectural decisions degrade or adapt as new features and constraints are injected into the system.

The five test scenarios are defined as follows:

\begin{enumerate}
\item \textbf{Scenario A: Game Development (Crossy Road).} Focuses on the evolution of game logic, starting from a basic prototype and advancing to a complex system with difficulty scaling, menu navigation, and an in-game currency economy.
\item \textbf{Scenario B: Utility Tool (Flashcard App).} Tracks the transition of a simple study tool into a robust GUI application featuring user profiles, theming (light/dark modes), and configurable spaced-repetition algorithms.
\item \textbf{Scenario C: Arcade Logic (Pacman).} Evaluates the agent's ability to implement collision detection, AI pathfinding, and state management within an Object-Oriented framework.
\item \textbf{Scenario D: Financial System (Personal Finance).} A high-complexity scenario requiring a dual-interface application (CLI and Web Dashboard) for budget tracking, data persistence, and visualization.
\item \textbf{Scenario E: Enterprise Backend (Online Bookstore).} The most complex test case, requiring the simulation of a full backend system with relational entities (books, authors, customers), inventory management, and sales reporting.
\end{enumerate}

\paragraph{Prompt Evolution Example}
To illustrate the incremental specification strategy, the four developmental stages for \textit{Scenario A} are detailed below. Note how requirements for logic, UI, and data structures are layered sequentially:

\begin{itemize}
    \item \textbf{Stage 1 (Base):} make a crossy road game in python using pygame. Use OOP paradigm, modularize the objects and files
    \item \textbf{Stage 2 (+Logic):} make a crossy road game in python using pygame. Make it become gradually harder. In the beginning it must be very easy. Use OOP paradigm, modularize the project

    \item \textbf{Stage 3 (+UI/State):} make a crossy road game in python using pygame. Make it become gradually harder. In the beginning it must be very easy. I also want a menu that shows different levels, so it's not one infinite loop game. Use OOP paradigm, modularize the project
    
    \item \textbf{Stage 4 (+Economy/Complex Integration):} make a crossy road game in python using pygame. Make it become gradually harder. In the beginning it must be very easy and slow. I also want a menu that shows different levels, so it's not one infinite loop game. Add coins to the game that you can collect, and later use them in menu to buy different skins/colors. Use OOP paradigm, modularize the project
\end{itemize}

By diversifying the domain complexity (Games vs. Utilities vs. Systems) and applying this graduated prompting pressure, we successfully leveraged MetaGPT to generate full-scale software projects with intricate directory structures. The full set of prompts for each scenario can be found in the publicly available replication package~\cite{ReplicationPackage}. 

\subsubsection{Code Smell Detection and Metric Collection}
To maintain methodological consistency, we employed the same static analysis pipeline as Experiment I, utilizing PyExamine~\cite{shivashankar2025pyexamine} to identify code, structural and architectural smells within the MetaGPT-generated repositories. However, given the macroscopic nature of Experiment II (full-scale projects versus isolated scripts), we expanded our data collection to include structural and process metrics.

For each generated application, we recorded the \textbf{Total Number of Files (ToF)} and the \textbf{Total Lines of Code (TLoC)} to quantify project size and complexity. TLoC measurements were standardized using the open-source utility \textit{cloc}~\cite{adanial_cloc}. Additionally, to correlate code quality with the effort exerted by the multi-agent system, we logged the \textbf{Agent Action Count (AAC)}—defined as the total number of operational steps executed by MetaGPT to fulfill the prompt requirements. Collecting these metrics allows us to normalize code smell frequencies against project size, enabling a granular analysis of smell density across varying levels of architectural complexity.

In summary, this dual-experiment design provides a holistic view of the coding capabilities of modern AI systems. By synthesizing the results from the isolated, algorithmic tasks of Experiment I and the complex, architectural challenges of Experiment II, we construct a comprehensive dataset of agent-generated technical debt. This data enables us to move beyond simple correctness metrics and rigorously evaluate the maintainability of AI-written code. The following section, Results and Discussion, presents our findings, detailing the specific patterns of code smells identified across the different models, prompting strategies, and architectural scales.

\section{Result and Discussion}
In this section, we present a detailed analysis of the code quality and architectural integrity exhibited by autonomous coding agents. We organize our findings to directly address the research questions posed in the introduction. First, we examine the baseline performance of LLMs on isolated algorithmic tasks (RQ1), quantifying the density of code smells under varying prompting conditions. Second, we investigate how the transition to complex, multi-agent system generation impacts architectural health (RQ2), identifying specific degradation patterns in larger codebases. Finally, we construct a taxonomy of the most prevalent smells observed across both experiments (RQ3), offering insights into the inherent 'coding habits' of current AI models. The section concludes with a discussion on the implications of these findings for the future design of self-correcting software agents.

\subsection{Result From Experiment I}
The count of detected code smells is detailed in Table~\ref{table: ex1-result-raw} for the Zero-Shot condition and in Table~\ref{table: ex1-result-np} for the Few-Shot condition.

\begin{table*}[hbt]
\centering
\caption{A comparison of code smell frequencies from the Zero-Shot experiment. The table highlights a fundamental difference between AI and human-generated code. While metrics like Feature Envy (FE) and Potential Shotgun Surgery (PSS) were found to be unreliable in this algorithmic context, two patterns stand out: a high incidence of Long Method (LM) smells in LLM-generated code (e.g., 11 instances for Qwen-Coder-480b) and the exclusive appearance of Temporal Field (TF) smells (4 instances) in the human baseline.}
\begin{tabular}{|l | l |l |l |l |l |l |l |l |l |}
\hline 
\textbf{Model} & \textbf{LM} & \textbf{FE} & \textbf{PSS} & \textbf{EC} & \textbf{TF}& \textbf{HCC} & \textbf{HFI}& \textbf{LF} & \textbf{TMB} \\
\hline
deepseek-coder-v2(16b)    & 2 & 6 & 0 & 0 & 0& 0& 2& 0&0\\\hline
Llama3 (70b)    & 0 & 1 & 0 & 0 & 0& 0 & 0 & 0& 0 \\\hline
Qwen Coder(30b)  & 7 & 3 & 2 & 3 & 0& 0& 0& 2& 0 \\\hline
Qwen Coder(480b)    & 11 & 9 & 1 & 0 & 0  & 0 & 2 & 0& 1 \\\hline
Gemini-2.5-pro    & 5 & 6 & 0 & 0& 0& 1& 1& 0& 1 \\\hline
Human Baseline    & 1 & 4 & 0 & 0& 4& 0& 1& 0& 1 \\\hline
\end{tabular}
\label{table: ex1-result-raw}
\end{table*}

\begin{table*}[hbt]
\centering
\caption{Raw counts of detected code smells for each model under the Few-Shot prompting condition. The data indicates that the fundamental defect patterns observed in the Zero-Shot experiment persist. The primary conclusion is that the few-shot examples failed to reduce the generation of Long Method (LM) smells, demonstrating the limitations of prompting for enforcing code quality.}
\begin{tabular}{|l | l |l |l |l |l |l |l |l |l |}
\hline 
\textbf{Model} & \textbf{LM} & \textbf{FE} & \textbf{PSS} & \textbf{EC} & \textbf{TF}& \textbf{HCC} & \textbf{HFI}& \textbf{LF} & \textbf{TMB} \\
\hline
deepseek-coder-v2(16b)    & 2 & 3 & 0 & 0 & 0& 0& 1 & 0& 0\\\hline
Llama3 (70b)    & 0 & 2 & 2 & 0 & 0& 0 & 0 & 0& 1 \\\hline
Qwen Coder(30b)  & 7 & 3 & 0 & 1 & 0& 0& 0& 0& 0 \\\hline
Qwen Coder(480b)    & 13 & 4 & 1 & 0 & 0  & 0 & 2 & 0& 1 \\\hline
Gemini-2.5-pro    & 8 & 7 & 0 & 0& 0& 0& 1& 0& 0 \\\hline
Human Baseline    & 1 & 4 & 0 & 0& 4& 0& 1& 0& 1 \\\hline
\end{tabular}
\label{table: ex1-result-np}
\end{table*}

The detection frequencies presented in Tables ~\ref{table: ex1-result-raw} and ~\ref{table: ex1-result-np} reveal that LLMs and human developers introduce distinct patterns of code smells. We analyze these findings by examining the validity of our analysis tool, comparing human and AI-generated code, and evaluating the impact of model scale.

\subsubsection{Validity of Static Analysis in Algorithmic Contexts}

Before analyzing aggregate trends, we manually audited flagged instances to validate the precision of PyExamine for single-file algorithmic solutions.

Threshold-based metrics—specifically Long Method (LM) and Long File (LF)—demonstrated high precision, correctly identifying functions and files that violated standard length recommendations.

However, we identified a systematic rate of False Positives (FP) for Feature Envy (FE) and Potential Shotgun Surgery (PSS) in this experimental context.

\begin{enumerate}
\item \textbf{Feature Envy (FE):} PyExamine flags functions making frequent calls to an external object. In our algorithmic setting, where helper objects (e.g., graph structures) are locally instantiated, the tool misinterpreted legitimate method chaining on these local variables as improper coupling.
\item \textbf{Potential Shotgun Surgery (PSS):} The tool flags high-frequency usage of a callable as a PSS risk. It flagged the native Python \texttt{range()} function, which is fundamental to iteration in algorithmic code, creating a false positive. While this heuristic is effective for detecting high coupling in large systems, it is not applicable here.
\end{enumerate}

Consequently, our analysis prioritizes all metrics except FE and PSS, as they provide the most reliable indicators of maintainability in this context.

\subsubsection{Distinct "Fingerprints" of Technical Debt}

Contrary to recent findings suggesting LLMs produce cleaner code~\cite{santa2025llm}, our results indicate that AI agents do not eliminate technical debt but rather shift its form.

A clear dichotomy emerged: human authors showed a propensity for state management smells, whereas LLMs introduced procedural ones. The human baseline produced 4 instances of Temporal Field (TF)—a pattern entirely absent in LLM outputs. A TF occurs when a class-level variable is used by only a subset of methods, reflecting a common practice in competitive programming to define variables globally for quick state management at the cost of proper encapsulation.

In contrast, LLMs—particularly the Qwen series and Gemini-2.5-pro were prone to the Long Method (LM) smell. Qwen-coder-480B, for instance, generated 11 LMs in a zero-shot setting, while the human baseline produced only one. This reveals a fundamental trade-off: humans tend to fragment state (TF) while keeping methods concise, whereas LLMs encapsulate state correctly but often fail to decompose complex procedures into smaller helper functions.

\subsubsection{The Inefficacy of Few-Shot Prompting for Decomposition}

A comparison of Table ~\ref{table: ex1-result-raw} and Table ~\ref{table: ex1-result-np} reveals that structured few-shot prompting yielded no meaningful improvement in code hygiene.

After excluding noise from FE and PSS metrics, the prevalence of LM remained stagnant or even worsened. For Qwen-coder-480B, LMs increased from 11 (Zero-Shot) to 13 (Few-Shot), and for Gemini-2.5-pro, they rose from 5 to 8.

We hypothesize that this is because it is difficult to demonstrate the absence of smells through static examples. Without explicit negative constraints (e.g., "Do not write functions longer than 20 lines"), the models may interpret the additional context in the prompt as a signal to generate more comprehensive—and thus more verbose—solutions, inadvertently exacerbating the LM smell.

\subsubsection{A Shift From State To Procedural Defects}
The analysis above confirms that LLMs do not eliminate technical debt but instead introduce a distinct "machine signature" of defects that differs fundamentally from that of human developers.
A clear dichotomy exists between human and agent-generated code. The human baseline frequently exhibited Temporal Fields (TF), indicating issues with state encapsulation. In contrast, code-generating agents showed a strong propensity for the Long Method (LM) smell. This tendency scales with model size; for instance, Qwen-Coder-480b generated 11 LMs in the zero-shot setting, while the human baseline produced only one.

\subsubsection{The Complexity of Failure in Large Models}

Our analysis highlights a nuanced relationship between model scale and the LM smell. When coding tasks proved too difficult for all models, the structural characteristics of their failed attempts diverged significantly.

Larger models like Qwen-coder-480B and Gemini-2.5-pro exhibited a higher density of LMs in their non-functional solutions. Manual inspection reveals this stems from their superior reasoning capacity; even in failure, they attempt to construct a comprehensive solution with extensive logic for edge cases, resulting in lengthy function bodies. In contrast, smaller models like Llama-3-70b tended to generate concise but fundamentally inadequate code. Thus, the presence of LMs in large models can serve as a proxy for attempted reasoning, while their absence in smaller models may simply indicate a failure to engage with the problem's complexity.

\subsubsection{The Reasoning-Complexity Paradox}
We observed a counter-intuitive relationship where greater model capability did not correlate with cleaner code. Instead, it correlated with method bloat.
\begin{itemize}
\item \textbf{High-Capacity Models:} Large models (e.g., Qwen-480b, Gemini-2.5-pro) attempt to handle complex edge cases with extensive procedural logic, often consolidating it into a single block and triggering LM alerts.
\item \textbf{Low-Capacity Models:} Smaller models often fail functionally before generating enough code to violate length-based smell thresholds.
\end{itemize}

Therefore, in this context, high code density in LLM-generated solutions is not just a flaw but can also be a proxy for the model's attempt at rigorous reasoning, albeit at the cost of readability and modularity.

\begin{conclusionbox}
\textbf{Findings For RQ1: A Shift in Technical Debt}
Our analysis reveals a fundamental dichotomy in technical debt. Human developers introduce state-management defects (e.g., Temporal Fields), while AI agents shift from state to procedural defects`` and exhibit a ``Reasoning-Complexity Paradox'': more capable models introduce procedural bloat (e.g., Long Method) as they attempt to handle complex logic, resulting in denser, less maintainable code.
\end{conclusionbox}

\subsection{Result From Experiment II}
\subsubsection{Shift in Code Smell Taxonomy}
\begin{figure*}
    \centering
    \includegraphics[width=0.65\textwidth]{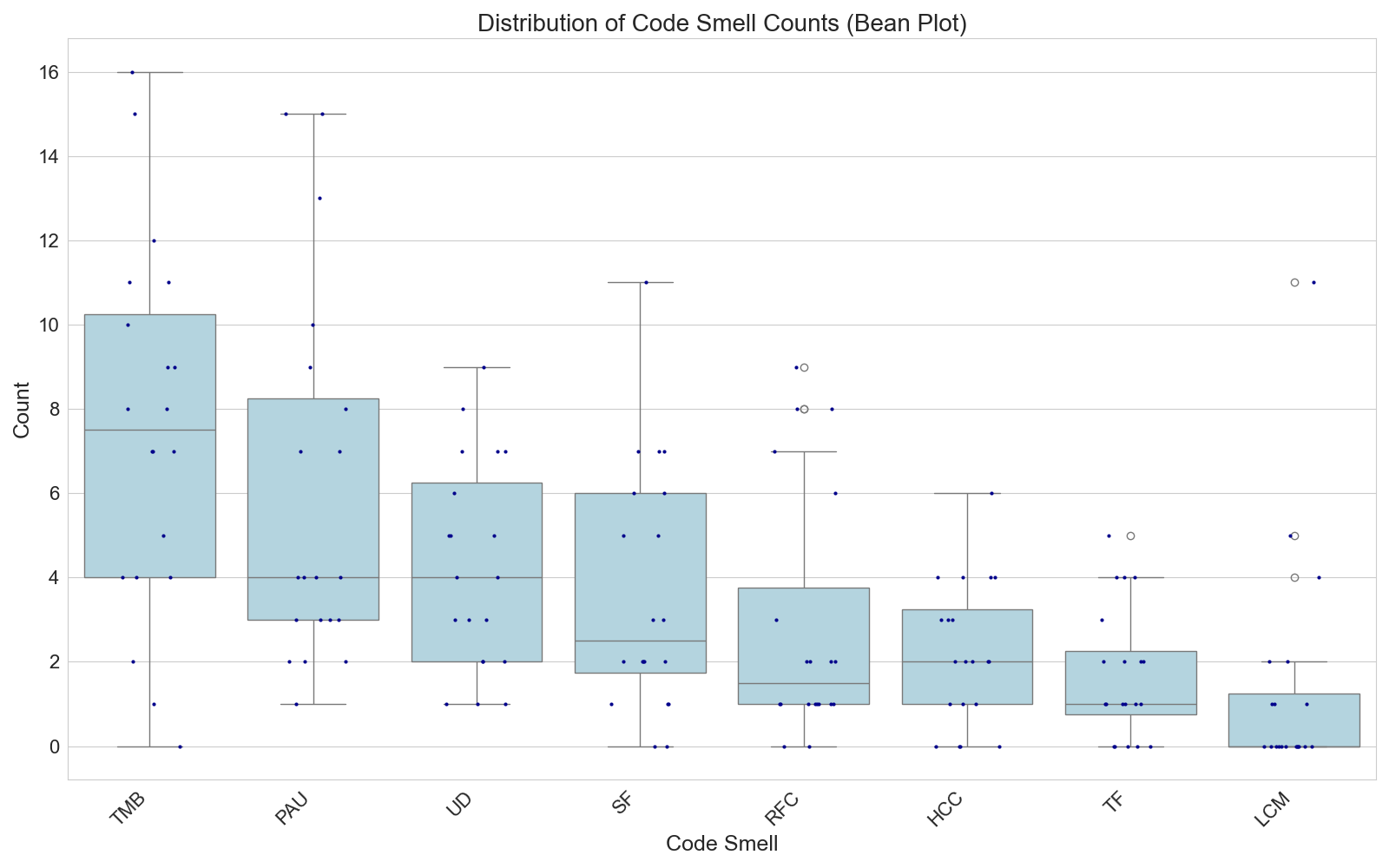}
    \caption{Distribution of Code Smell Counts. This box plot illustrates the distribution of counts for the most prevalent code smells, sorted in descending order by their mean value. Each box represents the interquartile range (IQR), with the central line denoting the median and the whiskers extending to 1.5 times the IQR. Points beyond the whiskers are plotted as individual outliers. The abbreviations for the code smells are as follows: TMB (Too Many Branches), PAU (Potential Improper API Usage), UD (Unstable Dependency), SF (Scattered Functionality), RFC (High Response for a Class), HCC (High Cyclomatic Complexity), TF (Temporal Field), and LCM (High Lack of Cohesion of Methods).}
    \Description{Box plot of the code smells introduced by MetaGPT.}
    \label{fig:five_plots}
\end{figure*}

In Figure \ref{fig:five_plots}, we plot the distribution of significant code smells detected in Experiment II.

In contrast to Experiment I, where algorithmic tasks primarily induced Long Method (LM) smells, the complex system generation in Experiment II triggers a distinct shift towards structural and architectural defects. As illustrated in Figure~\ref{fig:five_plots}, the most prevalent smells are now Too Many Branches (TMB), High Response for a Class (RFC), Potential Improper API Usage (PAU), Scattered Functionality (SF), and Unstable Dependencies (UD). Their consistent appearance across all stages and scenarios points to systemic limitations in the agent's design capabilities, not isolated errors.

\subsubsection{The "God Class" Syndrome (TMB \& RFC)}
The most prevalent smells were Too Many Branches (TMB) and High Response for a Class (RFC), indicating a systemic failure in responsibility distribution. The agent tends to centralize complex logic into singular "manager" classes rather than delegating behavior, resulting in brittle, tightly coupled components that are difficult to test.

\subsubsection{Redundant Implementation (PAU)}
Potential Improper API Usage (PAU) appears frequently, revealing a critical lack of local abstraction. Instead of encapsulating external interactions into reusable helper methods, the agent repeatedly rewrites invocation logic inline, mirroring a "copy-paste" coding style that inflates code volume and complicates maintenance.

\subsubsection{The Modular Mirage (SF \& UD)}
The co-occurrence of Scattered Functionality (SF) and Unstable Dependencies (UD) highlights a key limitation in agent reasoning. While the agent successfully splits code into multiple files (structural modularity), it fails to achieve semantic cohesion. Related behaviors become fragmented across the codebase, creating a "modular mirage" where file separation does not equate to logical separation.

These findings demonstrate that current agents achieve modularity only at a superficial, structural level. The dominance of these smells confirms that quality limitations stem from how agents organize responsibilities and control flow, not merely from how they write syntax. Future LLM-based systems must therefore incorporate explicit architectural constraints to mitigate these deficiencies.

\begin{conclusionbox}
\subsubsection*{Finding for RQ2: A Shift to Architectural Rot}
As tasks scale from algorithms to systems, the agent's defect profile shifts from procedural bloat to architectural decay. Our analysis reveals that the \textbf{top 5} most frequent defects are now structural and architectural smells, including "God Classes" (TMB \& RFC) and the "Modular Mirage" (SF \& UD), indicating a systemic failure to manage complexity.
\end{conclusionbox}

\subsection{Identifying the Drivers of Architectural Decay}

To validate our qualitative observations, we performed a statistical analysis to identify the key factors driving quality degradation. We measured code complexity using the Total Number of Files (ToF) and Total Lines of Code (TLoC), and categorized defects into code-, structural-, and architectural-level smells.

Given our sample size, we employed non-parametric statistical methods to identify key factors to predict existence of code smells. Given our relatively small sample size, the assumptions underlying traditional parametric tests—such as normal distribution and homogeneity of variance—are frequently violated and difficult to robustly verify. To mitigate the risk of Type I errors and ensure the validity of our findings, we utilized rank-based, distribution-free techniques. We used Spearman's Rank Correlation ($\rho$) to assess monotonic relationships between continuous variables (e.g., TLoC and smell counts) and the Kruskal-Wallis H-test for categorical factors (e.g., Scenario and Stage). The results are presented in Table~\ref{tab:stat_analysis}.

\begin{table}[t]
\centering
\caption{Statistical Analysis of Factors Affecting Code Smell Counts. \\ 
For Categorical factors (Scenario, Stage), values represent Kruskal-Wallis $H$ statistic ($p$-value). \\ 
For Continuous factors, values represent Spearman's Rank Correlation $\rho$ ($p$-value). \\
\textbf{Bold} values indicate statistical significance ($p < 0.05$).}
\label{tab:stat_analysis}
\begin{tabular}{lrrr}
\toprule
\textbf{Factor} & \textbf{Code} & \textbf{Structural} & \textbf{Architectural} \\
\midrule
\multicolumn{4}{l}{\textit{Categorical Factors (Kruskal-Wallis H-test)}} \\
\midrule
Scenario & 6.83 (0.145) & 8.51 (0.075) & \textbf{9.86 (0.043)*} \\
Stage    & 0.33 (0.955) & 0.62 (0.892) & 0.84 (0.840) \\
\midrule
\multicolumn{4}{l}{\textit{Continuous Factors (Spearman Correlation $\rho$)}} \\
\midrule
ToF      & 0.33 (0.153) & 0.06 (0.804) & \textbf{0.72 ($<$0.001)*} \\
TLoC     & \textbf{0.53 (0.017)*} & \textbf{0.59 (0.007)*} & \textbf{0.94 ($<$0.001)*} \\
ALpF & 0.23 (0.329) & \textbf{0.57 (0.009)*} & 0.30 (0.201) \\
\bottomrule
\end{tabular}
\end{table}

\subsubsection{The Volume-Quality Inverse Law}
Our analysis revealed that code volume is the single strongest predictor of quality degradation. \textbf{Total Lines of Code (TLoC)} showed a moderate positive correlation with code-level smells ($\rho=0.53$, $p=0.017$), which strengthened for structural smells ($\rho=0.59$, $p=0.007$). Most significantly, TLoC exhibited a near-perfect positive correlation with architectural smells ($\rho=0.94$, $p<0.001$). The \textbf{Total Number of Files (ToF)} was also a strong predictor of architectural decay ($\rho=0.72$, $p<0.001$). These findings establish a "Volume-Quality Inverse Law": as the agent generates more code and files, its ability to maintain a coherent architectural vision collapses.

\subsubsection{The Impact of Domain and Prompt Specificity}
The application \textbf{domain (Scenario)} also significantly impacted architectural quality ($H=9.86$, $p=0.043$), confirming that certain problem types are inherently more prone to architectural failure when handled by the agent. Counter-intuitively, the \textbf{specificity of requirements (Stage)} had no statistical impact on any smell category ($p > 0.8$). This critical finding implies that simply providing more detailed prompts is insufficient to mitigate architectural decay, pointing to a bottleneck in the agent's internal reasoning capacity rather than the external specification.

\begin{conclusionbox}
\subsubsection*{Finding for RQ3: Volume is the Primary Driver of Architectural Decay}
Architectural complexity is the primary driver of quality degradation, governed by a ``Volume-Quality Inverse Law.'' We found a near-perfect correlation between Total Lines of Code and architectural smells ($\rho=0.94$), while observing that more detailed prompts had no statistical effect on quality ($p>0.8$).
\end{conclusionbox}

\section{Threats to Validity}
We identify several potential threats to the validity of our study and discuss the mitigation strategies employed to address them.

\subsection{Internal Validity}
A potential limitation in Experiment I is the absence of an iterative "validate and correct" execution loop. Due to computational budget constraints, we evaluated the code quality of the initial solutions generated by the LLMs without providing them an opportunity to debug functional errors based on test case feedback. It is possible that an iterative process could lead to structural changes in the code. However, we argue that this design decision does not compromise our core conclusions. Our research objective is to characterize the \textit{inherent} coding habits and structural tendencies of the models. The presence of code smells—such as \textit{Long Method} or \textit{Temporal Fields}—in the initial output serves as a direct indicator of the model's training bias and architectural capability, regardless of whether the solution passes all functional test cases immediately.

\subsection{External Validity}
\textbf{Dataset Size and Diversity:} In Experiment I, we utilized a stratified sample of 90 problems from the CodeContest dataset rather than the complete corpus. While we ensured a balanced distribution across difficulty levels (Easy, Medium, Hard), a smaller sample size may not fully capture the long-tail distribution of algorithmic edge cases. Consequently, there may be specific problem types where model performance diverges from our observed trends. We intend to scale this analysis to the full dataset in future work to verify the statistical significance of these findings.

\textbf{Generalizability of Agent Frameworks:} Experiment II focuses exclusively on \textbf{MetaGPT} as the representative multi-agent framework, because it is a state-of-the-art fully automated code generating AI system at the time we conduct this research. The landscape of AI software generation is rapidly evolving with the emergence of new tools such as \textit{SpecKit}~\cite{Delimarsky2025spec} and \textit{Cursor Composer}. These systems may employ different orchestration logic or context-management strategies that could influence the architectural quality of the generated software. Therefore, our findings regarding architectural degradation in multi-agent systems should be interpreted as specific to the MetaGPT paradigm, though likely indicative of broader challenges in current agentic workflows. Extending this benchmark to cover these emerging platforms remains a critical direction for future research.

\section{Conclusion and Remarks}

This study investigated the structural quality of AI-generated software, moving beyond functional correctness to audit the technical debt introduced by both single LLMs and autonomous multi-agent systems. Our empirical analysis reveals that AI agents do not eliminate code smells but rather introduce a distinct ``machine signature'' of defects that evolves with task complexity. In simple algorithmic tasks, we identified a \textbf{Reasoning-Complexity Trade-off}, where more capable models generate bloated procedural logic (\textit{Long Method}) in their pursuit of comprehensive solutions. As complexity scales to multi-file system generation, this defect pattern shifts to severe architectural flaws, including ``God Classes'' and a ``Modular Mirage''---a superficial structural modularity that lacks true semantic cohesion.

We established a \textbf{Volume-Quality Inverse Law}, demonstrating that code volume is a near-perfect predictor of architectural decay, a trend that better prompting fails to mitigate. These findings lead us to conclude that current agents operate as proficient ``junior developers'': they follow instructions with syntactic precision but lack the ``senior architect'' foresight needed to manage system-wide dependencies and maintain a coherent architectural vision.

\subsection{Remarks on the Study}

The primary strength of this paper lies in its dual-experiment methodology, which allows for a multi-scale analysis of technical debt. By contrasting isolated algorithmic tasks with complex system generation, we successfully isolated how the nature of AI-generated defects shifts with architectural complexity. Furthermore, our focus on a well-defined taxonomy of code smells, rather than aggregate quality scores, provides a granular and interpretable view of the specific design ``habits'' of autonomous agents.

However, our study has limitations. The analysis in Experiment I was conducted without an iterative debugging loop, meaning the results reflect the models' inherent biases at first-pass generation. While this design choice was intentional to capture initial coding tendencies, it does not account for the potential of agents to self-correct structural flaws when provided with test feedback. In Experiment II, our findings are based on a single multi-agent framework (MetaGPT). Although MetaGPT\cite{hong2024metagpt} is a state-of-the-art representative at the time we conduct the research, the architectural patterns we observed may be specific to its Standard-Operating-Procedure-based paradigm. The rapidly evolving landscape of agentic workflows warrants caution when generalizing these findings to all multi-agent systems.

\subsection{Future Research Directions}

Our findings open several critical avenues for future research. The central challenge identified is no longer code generation but automated complexity management. We encourage the community to explore the following directions:

\begin{itemize}
    \item \textbf{Development of Architecturally-Aware Agents:} The inefficacy of prompting suggests a need for new agent architectures. Future systems could incorporate an explicit ``Architect'' agent role, responsible for defining and enforcing design constraints, or leverage techniques like constitutional AI to instill design principles directly into the model.

    \item \textbf{Automated Refactoring and Architectural Self-Healing:} Our work highlights the need for automated tools that can detect and remediate AI-specific smells. Research into ``Refactoring Agents'' that can identify issues like the ``Modular Mirage'' and propose large-scale refactoring operations would be a significant contribution.

    \item \textbf{Longitudinal Benchmarking of Software Evolution:} This study analyzed architectural decay across predefined stages. A valuable next step would be to create benchmarks that simulate continuous, open-ended software evolution, allowing researchers to study architectural rot over much longer timescales and under more dynamic conditions.

    \item \textbf{Metrics for Semantic Cohesion:} Our finding that agents produce structurally modular but semantically fragmented code underscores the need for new metrics. The development of automated measures that can quantify the semantic cohesion of a codebase would be invaluable for evaluating the true quality of AI-generated architectures.
\end{itemize}

Ultimately, bridging the gap between mere executability and long-term maintainability is essential for the successful integration of autonomous agents into real-world software engineering practices.

\section*{Acknowledgments}

Gemini 3.1 was used for editorial purposes.

\section*{Data Availability Statement}

To facilitate reproducibility for future researchers who wish to replicate or extend our results in this study, we provide the dataset and scripts in our replication package\cite{ReplicationPackage} at this url \url{https://doi.org/10.5281/zenodo.19245562}.
In this package, we include the Python scripts that download and preprocess the CodeContest dataset, as well as the scripts that replicate the experiment 1 with different LLMs. We also include the prompts for experiment II, as well as the script to process the generated code and report the results. A README file is provided in the package to instruct users how to use the package to replicate the experiments.

\bibliographystyle{ACM-Reference-Format}
\bibliography{references}

@misc{ReplicationPackage,
    author="Annonymous",
   title = "Replication Package", 
   howpublished = "\url{https://doi.org/10.5281/zenodo.19245562}",
   year = {2026}
  }

@article{he2025llm,
  title={LLM-Based Multi-Agent Systems for Software Engineering: Literature Review, Vision, and the Road Ahead},
  author={He, Junda and Treude, Christoph and Lo, David},
  journal={ACM Transactions on Software Engineering and Methodology},
  volume={34},
  number={5},
  pages={1--30},
  year={2025},
  publisher={ACM New York, NY}
}

@article{dong2025survey,
  title={A Survey on Code Generation with LLM-based Agents},
  author={Dong, Yihong and Jiang, Xue and Qian, Jiaru and Wang, Tian and Zhang, Kechi and Jin, Zhi and Li, Ge},
  journal={arXiv preprint arXiv:2508.00083},
  year={2025}
}

@article{Delimarsky2025spec,
  title   = "Diving Into Spec-Driven Development With GitHub Spec Kit",
  author  = "Delimarsky, Den",
  journal = "developer.microsoft.com/",
  year    = "2025",
  month   = "September",
  url     = "https://developer.microsoft.com/blog/spec-driven-development-spec-kit"
}

@article{li2022competition,
  title={Competition-Level Code Generation with AlphaCode},
    author={Li, Yujia and Choi, David and Chung, Junyoung and Kushman, Nate and
    Schrittwieser, Julian and Leblond, R{\'e}mi and Eccles, Tom and
    Keeling, James and Gimeno, Felix and Dal Lago, Agustin and
    Hubert, Thomas and Choy, Peter and de Masson d'Autume, Cyprien and
    Babuschkin, Igor and Chen, Xinyun and Huang, Po-Sen and Welbl, Johannes and
    Gowal, Sven and Cherepanov, Alexey and Molloy, James and
    Mankowitz, Daniel and Sutherland Robson, Esme and Kohli, Pushmeet and
    de Freitas, Nando and Kavukcuoglu, Koray and Vinyals, Oriol},
  journal={arXiv preprint arXiv:2203.07814},
  year={2022}
}

@inproceedings{hong2024metagpt,
  title={MetaGPT: Meta programming for a multi-agent collaborative framework},
  author={Hong, Sirui and Zhuge, Mingchen and Chen, Jonathan and Zheng, Xiawu and Cheng, Yuheng and Zhang, Ceyao and Wang, Jinlin and Wang, Zili and Yau, Steven Ka Shing and Lin, Zijuan and others},
  year={2024},
  organization={International Conference on Learning Representations, ICLR}
}

@inproceedings{batole2025llm,
  title={An LLM-Based Agent-Oriented Approach for Automated Code Design Issue Localization},
  author={Batole, Fraol and OBrien, David and Nguyen, Tien and Dyer, Robert and Rajan, Hridesh},
  booktitle={2025 IEEE/ACM 47th International Conference on Software Engineering (ICSE)},
  pages={637--637},
  year={2025},
  organization={IEEE Computer Society}
}

@article{naik2024enhancing,
  title={Enhancing Code Refactoring in Python: Leveraging Large Language Models},
  author={Naik, Arpita and SHYLAJA, RITHIKA RAJAN},
  Journal = {odr.chalmers.se},
  url = {https://odr.chalmers.se/server/api/core/bitstreams/85d14e4b-13dc-40c7-bcea-1fe230875b45/content},
  year={2024}
}

@article{wadhwa2024core,
  title={Core: Resolving code quality issues using llms},
  author={Wadhwa, Nalin and Pradhan, Jui and Sonwane, Atharv and Sahu, Surya Prakash and Natarajan, Nagarajan and Kanade, Aditya and Parthasarathy, Suresh and Rajamani, Sriram},
  journal={Proceedings of the ACM on Software Engineering},
  volume={1},
  number={FSE},
  pages={789--811},
  year={2024},
  publisher={ACM New York, NY, USA}
}

@article{cordeiro2024empirical,
  title={An empirical study on the code refactoring capability of large language models},
  author={Cordeiro, Jonathan and Noei, Shayan and Zou, Ying},
  journal={arXiv preprint arXiv:2411.02320},
  year={2024}
}

@article{liu2025exploring,
  title={Exploring the potential of general purpose LLMs in automated software refactoring: an empirical study},
  author={Liu, Bo and Jiang, Yanjie and Zhang, Yuxia and Niu, Nan and Li, Guangjie and Liu, Hui},
  journal={Automated Software Engineering},
  volume={32},
  number={1},
  pages={26},
  year={2025},
  publisher={Springer}
}

@article{siddeeq2025llm,
  title={LLM-based Multi-Agent System for Intelligent Refactoring of Haskell Code},
  author={Siddeeq, Shahbaz and Waseem, Muhammad and Rasheed, Zeeshan and Hasan, Md Mahade and Rasku, Jussi and Saari, Mika and Terho, Henri and Makela, Kalle and Kemell, Kai-Kristian and Abrahamsson, Pekka},
  journal={arXiv preprint arXiv:2506.19481},
  year={2025}
}

@inproceedings{pucho2025refactoring,
  title={Refactoring Python Code with LLM-Based Multi-Agent Systems: An Empirical Study in ML Software Projects},
  author={Pucho, Alexander Puma and Ferreira, Alexandre Mello and Cirilo, Elder Jos{\'e} Reioli and Cafeo, Bruno BP},
  booktitle={Simp{\'o}sio Brasileiro de Engenharia de Software (SBES)},
  pages={678--684},
  year={2025},
  organization={SBC}
}

@inproceedings{rajendran2025multi,
  title={A Multi-Agent LLM Environment for Software Design and Refactoring: A Conceptual Framework},
  author={Rajendran, Vasanth and Besiahgari, Dinesh and Patil, Sachin C and Chandrashekaraiah, Manjunath and Challagulla, Vishnu},
  booktitle={SoutheastCon 2025},
  pages={488--493},
  year={2025},
  organization={IEEE}
}

@article{xu2025mantra,
  title={Mantra: Enhancing automated method-level refactoring with contextual rag and multi-agent llm collaboration},
  author={Xu, Yisen and Lin, Feng and Yang, Jinqiu and Tsantalis, Nikolaos and others},
  journal={arXiv preprint arXiv:2503.14340},
  year={2025}
}

@inproceedings{shivashankar2025pyexamine,
  title={PyExamine: A Comprehensive, Un-Opinionated Smell Detection Tool for Python},
  author={Shivashankar, Karthik and Martini, Antonio},
  booktitle={2025 IEEE/ACM 22nd International Conference on Mining Software Repositories (MSR)},
  pages={763--774},
  year={2025},
  organization={IEEE}
}

@article{santa2025llm,
  title={Is LLM-Generated Code More Maintainable \& Reliable than Human-Written Code?},
  author={Santa Molison, Alfred and Moraes, Marcia and Melo, Glaucia and Santos, Fabio and Assun{\c{c}}ao, Wesley KG},
  journal={arXiv e-prints},
  pages={arXiv--2508},
  year={2025}
}

@article{chen2021evaluating,
  title={Evaluating large language models trained on code},
  author={Chen, Mark},
  journal={arXiv preprint arXiv:2107.03374},
  year={2021}
}

@article{zheng2023survey,
  title={A survey of large language models for code: Evolution, benchmarking, and future trends},
  author={Zheng, Zibin and Ning, Kaiwen and Wang, Yanlin and Zhang, Jingwen and Zheng, Dewu and Ye, Mingxi and Chen, Jiachi},
  journal={arXiv preprint arXiv:2311.10372},
  year={2023}
}

@inproceedings{zhang2023multilingual,
  title={Multilingual code co-evolution using large language models},
  author={Zhang, Jiyang and Nie, Pengyu and Li, Junyi Jessy and Gligoric, Milos},
  booktitle={Proceedings of the 31st ACM Joint European Software Engineering Conference and Symposium on the Foundations of Software Engineering},
  pages={695--707},
  year={2023}
}

@book{campbell2013sonarqube,
  title={SonarQube in action},
  author={Campbell, G Ann and Papapetrou, Patroklos P},
  year={2013},
  publisher={Manning Publications Co.}
}

@article{mo2019architecture,
  title={Architecture anti-patterns: Automatically detectable violations of design principles},
  author={Mo, Ran and Cai, Yuanfang and Kazman, Rick and Xiao, Lu and Feng, Qiong},
  journal={IEEE Transactions on Software Engineering},
  volume={47},
  number={5},
  pages={1008--1028},
  year={2019},
  publisher={IEEE}
}

@software{adanial_cloc,
  author       = {Albert Danial},
  title        = {cloc: v1.92},
  month        = dec,
  year         = 2021,
  publisher    = {Zenodo},
  version      = {v1.92},
  doi          = {10.5281/zenodo.5760077},
  url          = {https://doi.org/10.5281/zenodo.5760077}
}

\end{document}